\documentclass[onecolumn]{article}
\usepackage{graphicx}
\usepackage{caption2}
\usepackage{amssymb}
\usepackage{times}
\usepackage{emulateapj}
\usepackage{psfig}

\def\spose#1{\hbox to 0pt{#1\hss}}

\let\approxlt=\lesssim
\let\approxgt=\gtrsim

\def\multleft#1{\hbox to size{\vbox {\halign {\lft{##}\cr #1}}\hfill}\par}
\def\multright#1{\hbox to size{\vbox {\halign {\rt{##}\cr #1}}\hfill}\par}

\def\boxit#1{\vbox{\hrule\hbox{\vrule\kern3pt\vbox{\kern3pt
          #1 \kern3pt}\kern3pt\vrule}\hrule}}

\def\Msun{\hbox{$\rm\thinspace M_{\odot}$}}
\def\pc{{\rm\thinspace pc}}



\begin{document}

\title{Sporadically Torqued Accretion Disks Around Black Holes}

\author{David~Garofalo\altaffilmark{1} and 
Christopher~S.~Reynolds\altaffilmark{2}}
\altaffiltext{1}{Dept.of Physics, University of Maryland, College Park, MD20742}
\altaffiltext{2}{Dept.of Astronomy, University of Maryland, College Park, MD20742}

\begin{abstract}
  The assumption that black hole accretion disks possess an untorqued
  inner boundary, the so-called zero torque boundary condition, has
  been employed by models of black hole disks for many years.
  However, recent theoretical and observational work suggests that
  magnetic forces may appreciably torque the inner disk.  This raises
  the question of the effect that a time-changing magnetic torque may
  have on the evolution of such a disk.  In particular, we explore the
  suggestion that the ``Deep Minimum State'' of the Seyfert galaxy
  MCG--6-30-15 can be identified as a sporadic inner disk torquing
  event.  This suggestion is motivated by detailed analyses of changes
  in the profile of the broad fluorescence iron line in {\it
    XMM-Newton} spectra.  We find that the response of such a disk to
  a torquing event has two phases; an initial damming of the accretion
  flow together with a partial draining of the disk interior to the
  torque location, followed by a replenishment of the inner disk as
  the system achieves a new (torqued) steady-state.  If the Deep
  Minimum State of MCG--6-30-15 is indeed due to a sporadic torquing
  event, we show that the fraction of the dissipated energy going into
  X-rays must be smaller in the torqued state.  We propose one such
  scenario in which Compton cooling of the disk corona by ``returning
  radiation'' accompanying a central-torquing event suppresses the
  0.5--10\,keV X-ray flux coming from all but the innermost regions of
  the disk.
\end{abstract}


\section{Introduction}

There is direct evidence that active galactic nuclei (AGN) are powered
by disk accretion onto supermassive black holes.  For example, the
central regions of the giant elliptical galaxy M87 were spatially
resolved by the {\it Hubble Space Telescope} sufficiently to actually
see a $\sim 100\pc$-scale disk of ionized gas in orbit about an unseen
mass of $3\times 10^9\,M_\odot$, presumed to be a supermassive black
hole (Ford et al, 1994; Harms et al, 1994).  The fact that this gas
disk is approximately normal to the famous relativistic jet displayed
by this object is circumstantial evidence that it is, indeed, the
outer regions of the accretion disk that powers this AGN.  In another
important case, radio observations of megamasers in the spiral galaxy
NGC~4258 (M106) reveal an almost perfectly Keplerian $\sim 0.5\pc$ gas
disk orbiting a compact object with mass $3.5\times 10^7\,M_\odot$
(Miyoshi et al. 1995; Greenhill et al. 1995).

Even before the observational evidence for black hole accretion disks
became compelling, the basic theory of such disks had been extensively
developed.  Building upon the non-relativistic theory of Shakura \&
Sunyaev (1973), Novikov \& Thorne (1974) and Page \& Thorne (1974)
developed the ``standard'' model of a geometrically-thin,
radiatively-efficient, steady-state, viscous accretion disk around an
isolated Kerr black hole.  In addition to the assumptions already
listed, it is assumed that the viscous torque operating within the
disk becomes zero at the radius of marginal stability, $r=r_{\rm ms}$.
Physically, this was justified by assuming that the accretion flow
would pass through a sonic point close to $r=r_{\rm ms}$ and hence flow
ballistically (i.e., ``plunge'') into the black hole.  

Even while setting up this boundary condition, Page \& Thorne (1974)
noted that magnetic fields may allow this zero-torque boundary
condition (ZTBC) to be violated.  Given the modern viewpoint of
accretion disks, that the very ``viscosity'' driving accretion is due
to magnetohydrodynamic (MHD) turbulence, the idea that the ZTBC can be
violated has been revived by recent theoretical work, starting with
Gammie (1999) and Krolik (1999a).  In independent treatments, these
authors show that significant energy and angular momentum can be
extracted from matter within the radius of marginal stability via
magnetic connections with the main body of the accretion disk.  Agol
\& Krolik (2000) have performed the formal extension of the standard
model to include a torque at $r=r_{\rm ms}$ and show that the extra
dissipation associated with this torque produces a very centrally
concentrated dissipation profile.  As shown by Gammie (1999), Agol \&
Krolik (2000), and Li (2002), this process can lead to an extraction
(and subsequent dissipation) of spin energy and angular momentum from
the rotating black hole by the accretion disk.  In these cases, the
magnetic forces might be capable of placing the innermost part of the
flow on negative energy orbits, allowing a Penrose process to be
realized (we note that Williams [2003] has also argued for the
importance of a non-magnetic, particle-particle and particle-photon
scattering mediated Penrose process).  A second mechanism by which the
central accretion disk can be torqued is via a direct magnetic
connection between the inner accretion disk and the (rotating) event
horizon of the black hole.  In this case, as long as the angular
velocity of the event horizon exceeds that of the inner disk, energy
and angular momentum of the spinning black hole can be extracted via
the Blandford-Znajek mechanism (Blandford \& Znajek 1977).  We note
that field lines that directly connect the rotating event horizon with
the body of the accretion disk through the plunging region {\it are}
seen in recent General Relativistic MHD simulations of black hole
accretion (e.g., Hirose et al. 2004).

Interest in these torqued relativistic disks has received a boost from
recent X-ray observations.  The X-ray spectra of many AGN (and,
indeed, Galactic Black Hole Candidates) reveal the signatures of
``X-ray reflection'' from optically-thick matter.  In some cases,
examination of these features allows us to detect strong Doppler and
gravitational shifts indicative of circular motion close to a black
hole (Fabian et al. 1989; Tanaka et al. 1995; Fabian et al. 2000;
Reynolds \& Nowak 2003).  In these cases, the observed range of
Doppler and gravitational shifts can be used to map the X-ray
emission/irradiation across the surface of the accretion disk.  The
{\it XMM-Newton} satellite is particularly well suited to the study of
these features due to its combination of good spectral resolution and
high throughput.  Using {\it XMM-Newton}, Wilms et al. (2001) and
Reynolds et al. (2004) studied the Seyfert-1 galaxy MCG--6-30-15 in
its peculiar ``Deep Minimum State'' first discovered by ASCA (Iwasawa
et al. 1996).  Confirming the principal result of Iwasawa et al.
(1996), the X-ray reflection features were found to be extremely
broad.  The degree of gravitational redshifting required the majority
of the X-ray emission to emerge within a radius of $r\sim 2GM/c^2$
from a near-extremal Kerr black hole (i.e., a black hole with a spin
parameter of $a=0.998$).  As explicitly shown in Reynolds et al.
(2004), it is very problematic to explain these data within the
framework of the standard accretion disk model.  Fabian \& Vaughan
(2003) and Miniutti \& Fabian (2004) suggest that gravitational
focusing of the primary continuum X-rays might produce such a
centrally-concentrated emissivity profile.  Alternatively, Reynolds et
al. (2004) has shown that a torqued disk can readily explain the Deep
Minimum spectrum provided the source is assumed to be in a
torque-dominated state (or, in the terminology of Agol \& Krolik
[2000], an ``infinite-efficiency'' state) whereby the power associated
with the innermost torque is instantaneously dominating the accretion
power.  In other words, the X-ray data suggest that during this Deep
Minimum state of MCG--6-30-15 the power derived from the black hole
spin greatly exceeds that derived from accretion.

Of course, this state of affairs cannot last forever or else the
central black hole in MCG--6-30-15 would spin down to a point where it
could no longer provide this power.  At some point in its history, the
system must be in an accretion-dominated phase in which the black hole
is spun up.  However, even in its spin-dominated state, the spin-down
timescale of the central black hole is of the order of 100 million
years or more.  Thus we could envisage a situation in which the system
shines via a quasi-steady-state, spin-dominated accretion disk.  There
are hints, though, that accretion disks may switch between
spin-dominated and accretion-dominated on much shorter timescales.  In
its normal spectral state, the X-ray reflection features in
MCG--6-30-15 are much less centrally concentrated than in the Deep
Minimum State, suggesting that the normal state might be
accretion-dominated.  It is also important to note that this system
can switch between its normal state and the Deep Minimum State in as
little as 5--10\,ksec (Iwasawa et al.  1996), which corresponds to
only a few dynamical timescales of the inner accretion disk.  Thus it
is of interest to consider the physics of an accretion disk that
undergoes a rapid torquing event.  That is the prime motivation for
this paper.

In Section 2 we will begin our study of sporadically torqued accretion
disks by investigating an analytic solution for a torqued Newtonian
disk.  In Section 3, we generalize to the fully relativistic equations
and obtain numerical solutions.  In Section 4, we relate our results
with the observed properties of the ``Deep Minimum State'' of
MCG--6-30-15, and consequently discuss the effect that a torquing
event may have on the physics of the X-ray emitting disk corona.  In
particular, we suggest that the enhanced Returning Radiation
associated with a torquing event might suppress 0.5--10\,keV coronal
emission in all but the inner portion of the disk.  Section 5
summarizes our main conclusions.

\section{An analytic ``toy'' model of a torqued disk}

We begin our investigation of sporadically torqued disks via the study
of a simple case that lends itself to a straightforward analytic
solution.

We shall construct a model of a radiatively-efficient accretion disk
following the usual approach of Pringle (1981).  We shall assume that
the accretion disk is axisymmetric, geometrically-thin and in
Keplerian motion about a point-mass $M$.  Using a cylindrical polar
coordinate system $(r,z,\phi)$ with the axis passing through the
central mass normal to the disk plane, we shall denote the surface
density of the disk by $\Sigma(r,t)$, the angular velocity of the disk
about $M$ as $\Omega(r)$ and the radial velocity of the disk material
as $v_r(r,t)$.  The equations that determine the structure of the thin
disk assuming radiative efficiency are mass and angular momentum
conservation,
\begin{equation}
\label{eqn:mass_con}
r\frac{\partial\Sigma}{{\partial}t}+\frac{{\partial}(rv_{r}\Sigma)}{{\partial}r}=0,
\end{equation}
\begin{equation}
\label{eqn:ang_con}
r\frac{{\partial}(\Sigma r^2\Omega)}{{\partial}t}+\frac{{\partial}(r\Sigma v_rr^2\Omega)}{\partial r}=\frac{1}{2\pi}\frac{{\partial}G}{{\partial}r},
\end{equation}
respectively, where $G(r,t)$ is the torque exerted by the disk {\it
outside} of radius $r$ on the disk {\it inside} of that radius.

In standard disk models, the torque $G$ is the integrated value of the
only stress tensor component ($S_{r\phi}$) that survives the condition
of axisymmetry and geometric-thinness.  From Krolik (1999b) we have,
\begin{equation}
G\,=\,\int\ rS_{r\phi}\,dz \int\ r\,d\phi\,=\,2\pi r^{3}\nu\Sigma\,\frac {\partial\Omega}{\partial r},  
\end{equation}
where we have introduced an ''effective kinematic viscosity'', $\nu$.  To generalize these models to the case of an externally imposed torque, we set
\begin{equation}
\label{eqn:visc_def}
G=2\pi r^{3}\nu\Sigma\,\frac {\partial\Omega}{\partial r} + G_T.
\end{equation}
Combining eqns.~\ref{eqn:mass_con}, \ref{eqn:ang_con} and
\ref{eqn:visc_def}, and specializing to a Keplerian rotation curve, we
get the usual diffusion equation for surface density modified for the
effects of the external torque,
\begin{equation}
\frac{\partial\Sigma}{\partial t}=\frac{3}{r}\frac{\partial}{\partial r}\left[r^{1/2}\frac{\partial(\nu\Sigma r^{1/2})}{\partial r}\right]-\frac{1}{r\pi(GM)^{1/2}}\frac{\partial}{\partial r}\left(r^{1/2}\frac{\partial G_{T}}{\partial r}\right).
\label{eqn:basic_diffusion}
\end{equation}
For the rest of this paper, we shall work in units where $GM=1$.
Changing variables to $x=r^{1/2}$ and $\psi=\nu\Sigma x$ and assuming
that $\nu$ has no explicit time dependence, we get
\begin{equation}
\label{eqn:diffusion}
\frac{\partial \psi}{\partial t}=\frac{3\nu}{4x^2}\frac{\partial^2}{\partial x^2}\left(\psi - \frac{G_T}{3\pi}\right).
\end{equation}

We now consider a particular torquing event.  Suppose that the disk
suffers no external torques for the period $t<0$.  Then, at $t=0$, we
engage an external torque (possibly resulting from a magnetic
connection to the plunging region or spinning event horizon) that
deposits angular momentum into a narrow annulus at $r=r_0$.  If the
rate at which angular momentum is being deposited is $\beta$, we have
\begin{equation}
\frac{\partial G_T}{\partial r}=\beta \delta(r-r_0)\Theta(t),
\end{equation}
giving
\begin{equation}
G_T(r,t)=\beta \Theta(r-r_0)\Theta(t),
\end{equation}
where $\Theta$ is the Heavyside step function.  At this point, we specialize to a particular viscosity law.  We set $\nu=kr$ for mathematical convenience, although it will not make qualitative difference.  We can now rewrite
eqn.~\ref{eqn:diffusion} as,
\begin{equation}
\label{eqn:mod_diff}
\frac{\partial \xi}{\partial t}=\frac{3k}{4}\left[\frac{\partial^2 \xi}{\partial x^2}\right]-\frac{\beta}{3\pi}\Theta(x-x_0)\delta(t),
\end{equation}
where,
\begin{eqnarray}
\xi&=&\psi-\frac{\beta}{3\pi}\Theta(x-x_0)\Theta(t),
\end{eqnarray}
and the delta-function in time results from the time-derivative of the
$\Theta(t)$ term.

Suppose that the disk is in the untorqued steady state at $t<0$.  From
eqn.\ref{eqn:mod_diff}, one can easily see that such a steady state is
given by,
\begin{equation}
\xi_{ss}=\psi=A(x-x_i)\hspace{1cm}(t<0),
\end{equation}
where $A$ is a normalization constant and $r_i\equiv x_i^2$ is the
inner edge of the untorqued disk defined as the location where the
``viscous'' torques vanish.  Examination of eqn.~\ref{eqn:mod_diff}
shows that the time-dependent behavior of the torqued disk at times
$t>0$ is given by the simple diffusion equation,
\begin{equation}
\frac{\partial \xi}{\partial t}=\frac{3k}{4}\frac{\partial^2 \xi}{\partial x^2},
\end{equation}
with an initial condition set by integrating through the
delta-function in time, $\xi(x,t=0)=\xi_{ss}-\beta\Theta(x-x_0)/3\pi$.
The appropriate boundary condition is $\xi\rightarrow \xi_{ss}$ as
$t\rightarrow\infty$.  Standard methods (i.e., separation of
variables) give the following solution:
\begin{equation}
\xi(x,t)=\xi_{ss}(x)
+\frac{\beta}{3\pi^2}\int_0^\infty\frac{1}{\lambda}[\sin\lambda x_0\,\cos\lambda x\,-(\cos\lambda x_0\,+1)\sin\lambda x]\exp\left[-3k\lambda^2t/4\right]\,d\lambda.
\end{equation}

With this solution, we can compute the surface density of the disk at
any given radius and time.  Armed with the surface density, we can
then compute all other quantities of interest including the viscous
dissipation rate per unit surface area of the disk:
\begin{equation}
D(r)=\frac{\nu\Sigma r^2}{2}\Omega^{\prime 2},
\end{equation}
where $\Omega^\prime=d\Omega/dr$.  Plots of the radial dependence of
$\Sigma(r)$ and $D(r)$ for various times are shown in
Figures~1--\ref{fig:newt2}.  Also shown is the time dependence of the
total viscous dissipation obtained by integrating $D(r)$ across the
whole disk (i.e. luminosity).  We can see that the response of the
disk to the onset of an external torque can be separated into two
phases.  In the first phase, the accretion flow is ``dammed'' at
$r=r_0$ due to the inability of the accretion flow to transport the
angular momentum deposited by the external torque.  This leads to a
build-up of mass (i.e., an increase in the surface density) in the
region $r>r_0$.  Concurrently, matter in the region $r<r_0$ continues
to accrete thereby partially draining away the surface density.  The
inevitable result is a growing discontinuity in the surface density at
$r=r_0$.  The angular momentum transport associated with this
discontinuity grows until mass can, once again, flow inwards across
this radius.  One then enters the second phase of evolution, whereby
the surface density in the region $r<r_0$ is replenished back to its
original level while the surface density discontinuity is maintained
at approximately a constant level.  Eventually, one achieves the
torqued steady-state solution (e.g., Agol \& Krolik 2000).  The two
sets of figures are for an external torque at $r=4$ and one closer to
the inner edge at $r=2$.  Note how, for a given injection rate of
angular momentum, the effect on the disk structure is much more
dramatic for smaller radius.

Since our disks are assumed to be radiatively-efficient, the
instantaneous total luminosity of the accretion disk can be formally
decomposed into two components, one due to the decrease in
gravitational potential energy of the accreting gas, and a second due
to the work done by the external torque, i.e., 
\begin{equation}
L=2\int 2\pi rD(r)dr=\frac{1}{2}\int\;\frac{GM \dot{M}}{r^{2}}dr+\int \Omega \frac{\partial G_{T}}{\partial r}dr\normalsize.
\end{equation}   
As can be seen from Fig.~1(f) and Fig.~\ref{fig:newt2}(f), the
luminosity dips before climbing up to a new elevated level that
includes the work done by the external torque as well as the accretion
energy.  The temporary dip in luminosity is due to the damming of the
accretion flow in the early evolution of the torquing event.

\begin{figure*}
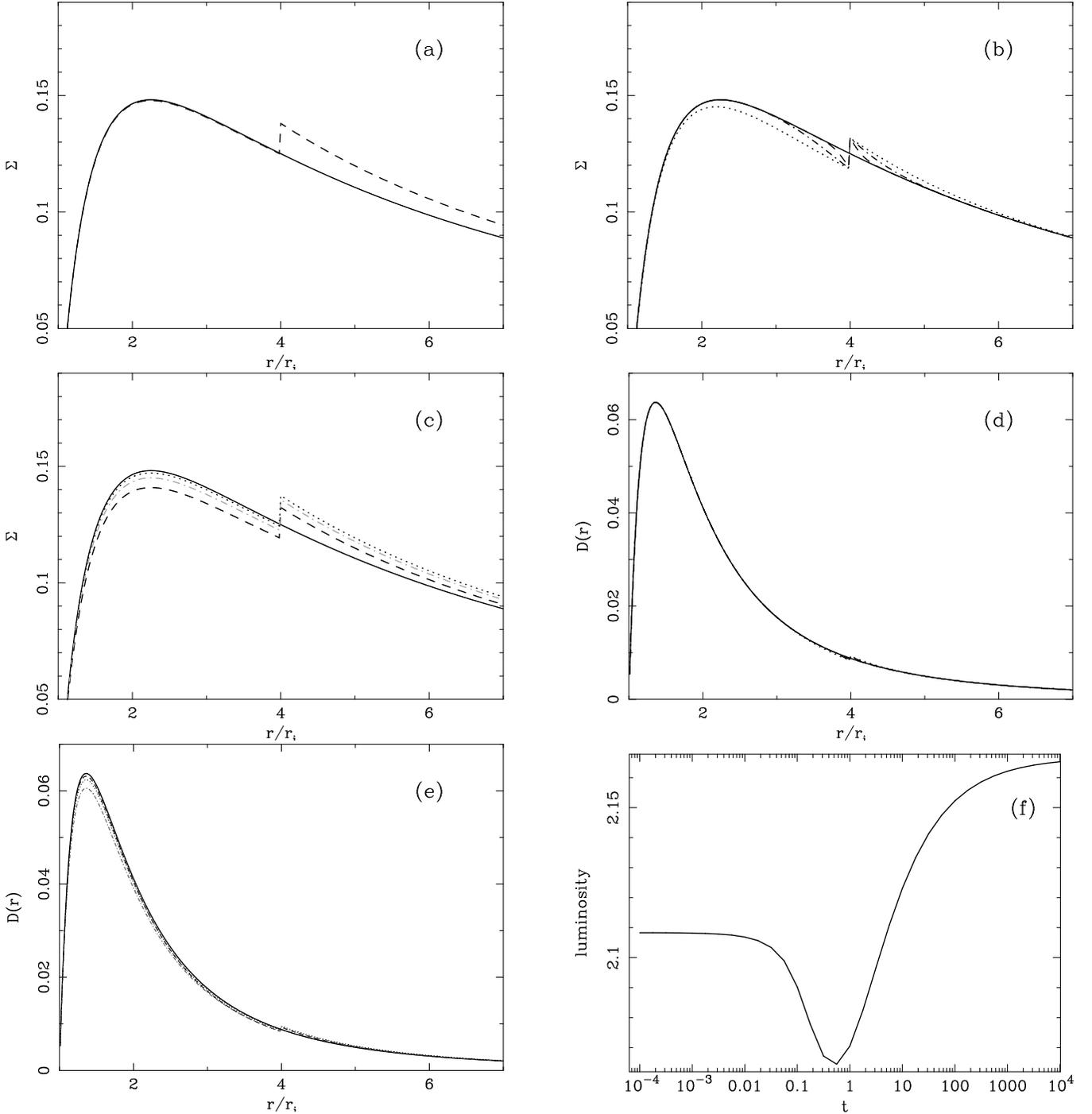

\hbox{ \psfig{figure=f1a.ps,width=0.45\textwidth,angle=270}
\hspace{1cm}
\psfig{figure=f1b.ps,width=0.45\textwidth,angle=270}
}
\hbox{
\psfig{figure=f1c.ps,width=0.45\textwidth,angle=270}
\hspace{1cm}
\psfig{figure=f1d.ps,width=0.45\textwidth,angle=270}
}
\hbox{
\psfig{figure=f1e.ps,width=0.45\textwidth,angle=270}
\hspace{1cm}
\psfig{figure=f1f.ps,width=0.45\textwidth,angle=270}
}
\label{one}
\caption{Evolution of our ``toy'' model disk with a torque acting at $r=4$.  
  Panel (a) shows the initial state of the surface density profile for
  the non torqued disk (solid line) and the resulting torqued
  steady-state (dashed line).  Panel (b) shows four times in the early
  evolution of the surface density profile (solid-line:$t=0$ the
  untorqued steady-state, dashed-line:$t=10^{-3}$,
  dot-dashed-line:$t=10^{-2}$, dotted-line:$t=10^{-1}$; we use units
  such that $k=1$ which corresponds to scaling with respect to the
  viscous timescale of the inner disk).  Notice how the initial
  evolution is such that density drops inward of the torque location
  and increases outward of it due to the ``damming'' of the accretion
  flow.  Panel (c) shows the subsequent late evolution towards the
  torqued steady-state (solid-line:$t=0$, or untorqued steady-state,
  dashed-line:$t=1$, dot-dashed-line:$t=10$, dotted-line:$t=100$).
  Panels (d) and (e) show the dissipation profiles $D(r)$ in the early
  and late stages, respectively, of the evolution with the type of
  line and time corresponding to those of figures (b) and (c).  In
  this case, the effect on $D(r)$ is subtle.  In panel (f), we show
  the luminosity profile obtained by integrating the dissipation
  profile over the disk surface.  The final steady-state torqued
  luminosity profile is enhanced with respect to the non-torqued
  steady-state profile due to work done by the torque.}
\end{figure*}

\begin{figure*}
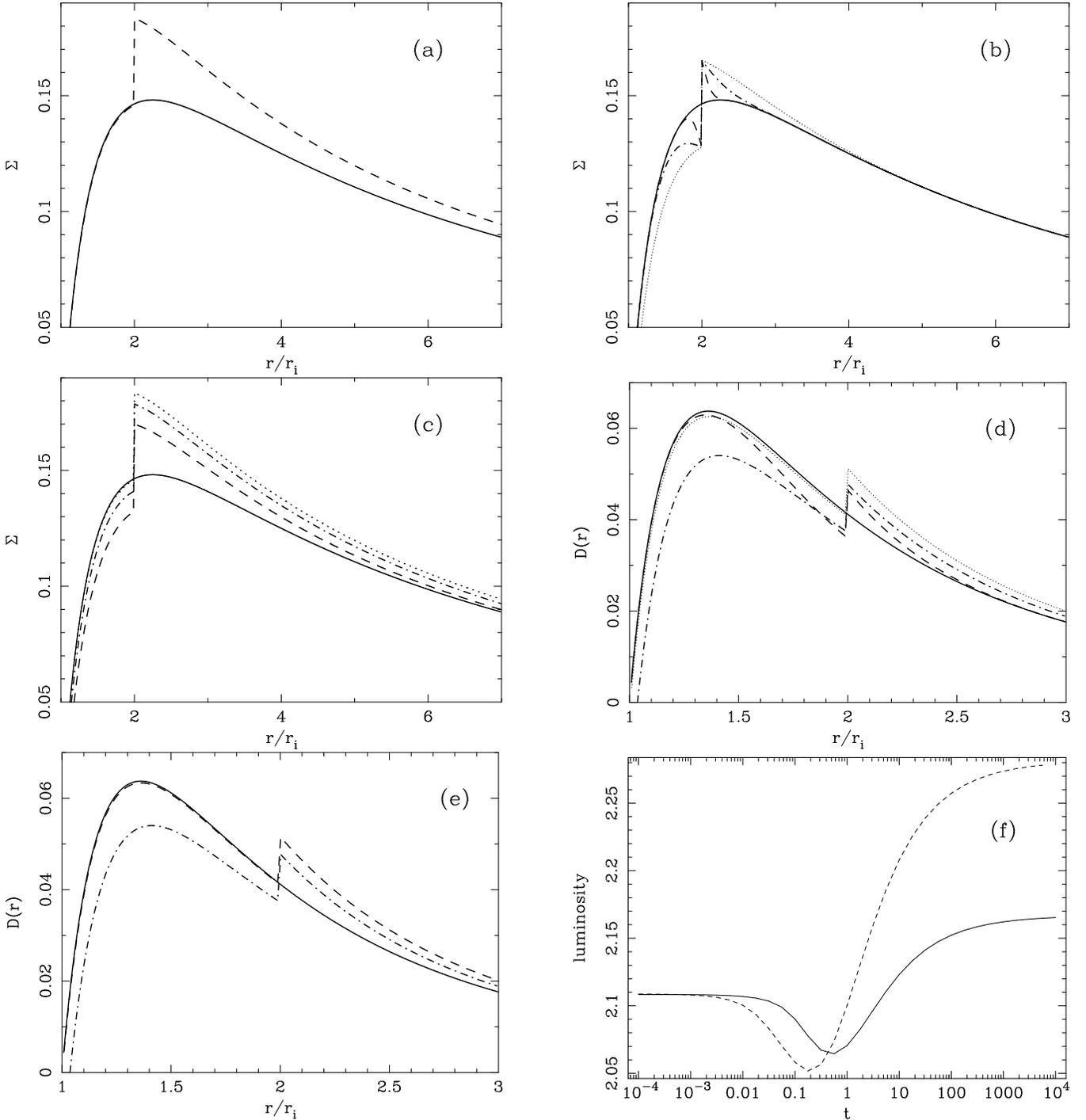

\hbox{
\psfig{figure=f2a.ps,width=0.45\textwidth,angle=270}
\hspace{1cm}
\psfig{figure=f2b.ps,width=0.45\textwidth,angle=270}
}
\hbox{
\psfig{figure=f2c.ps,width=0.45\textwidth,angle=270}
\hspace{1cm}
\psfig{figure=f2d.ps,width=0.45\textwidth,angle=270}
}
\hbox{
\psfig{figure=f2e.ps,width=0.45\textwidth,angle=270}
\hspace{1cm}
\psfig{figure=f2f.ps,width=0.45\textwidth,angle=270}
 }
\label{fig:newt2}
\caption{Evolution of our ``toy'' model disk with a torque acting at $r=2$.
  Panel (a) shows the initial state of the surface density profile for
  the non torqued disk (solid line) and the resulting torqued
  steady-state (dashed line).  Panel (b) shows four times in the early
  evolution of the surface density profile (solid-line:$t=0$,
  dashed-line:$t=10^{-3}$, dot-dashed-line:$t=10^{-2}$,
  dotted-line:$t=10^{-1}$).  Panel (c) shows the subsequent late
  evolution towards the torqued steady-state (dashed-line:$t=1$,
  dot-dashed-line:$t=10$, dotted-line:$t=100$).  For reference, the lower solid-line shows the
  untorqued steady-state disk.  Panels (d) and (e) show the
  dissipation profiles $D(r)$ in the early and late stages,
  respectively, of the evolution with the same line-type/time arrangement of figure 1.  In other words, the line types of figure (b) match those of figure (d) and those of figure (c) match those of figure (e).  In panel (f), we show the luminosity
  profile obtained by integrating the dissipation profile over the
  disk surface for both torques at r=2 (dashed line) and r=4 (solid line).}
\end{figure*}

Now that we have explored a torquing event via the analytical solution
of an extremely simplified accretion disk model, we move on to
somewhat more realistic models.  In the next section, we present a
semi-analytic analysis of a geometrically-thin general relativistic
accretion disk.

\section{Relativistic torqued accretion disks}

Relativity produces two complications to the analysis.  Firstly, the
equations governing the structure of the accretion disk are rather
more complex and drive us to use numerical rather than analytic
techniques.  Secondly, the relationship between the emitted and
observed fluxes becomes non-trivial, with gravitational light bending,
relativistic aberration/beaming, and Doppler/gravitational redshifting
all becoming important.  We shall deal with these issues in turn.

The time-dependent equations describing the structure of a
geometrically-thin accretion disk in the $\theta=\pi/2$ plane of a
Kerr spacetime are given in Boyer-Lindquist coordinates
$(t,R,\theta,\phi)$ by Eardley \& Lightman (1974).  Taking $\Sigma(R)$
to be the proper surface density of the disk (i.e., the surface
density measured by a local observer moving with the fluid), the disk
evolution is described by
\begin{equation}
\label{eqn:rel_diff}
\frac{\partial\Sigma}{{\partial}t}=\frac{{\cal C}^{1/2}}{{\cal B}R}\frac{\partial}{\partial R}\left[\frac{\Gamma}{\frac{\partial L^\dagger}{\partial R}}\frac{\partial}{\partial R}({\cal W}R^{2}{\cal D})\right],
\end{equation}
where,
\begin{eqnarray}
{\cal B}&=&(1+\frac{aM^{1/2}}{R^{3/2}}),\\
{\cal C}&=&(1-\frac{3M}{R}+\frac{2aM^{1/2}}{R^{3/2}}),\\
{\cal D}&=&(1-\frac{2M}{R}+\frac{a^{2}}{R^{2}}),\\
\Gamma&=&\frac{{\cal B}}{{\cal C}^{1/2}},\\
L^\dagger&=&M^{1/2}R^{1/2}(1-\frac{2aM^{3/2}}{R^{3/2}}+\frac{a^2M^2}{R^2}).
\end{eqnarray}
Here, $a$ is the dimensionless spin parameter of the black hole
(denoted as $a_*$ by Eardley \& Lightman 1974) and $L^\dagger$ is the
specific angular momentum of the fluid for prograde orbits.  The local
$r\phi$ shear in this flow is $\sigma=-3\Omega {\cal C}^{-1}{\cal D}$,
where $\Omega=(M/r^3)^{1/2}$.  Thus, guided by the non-relativistic
prescription, we set the vertically-integrated $r\phi$ component of
the stress tensor in the absence of an external torque to be
\begin{equation}
{\cal W}=-\nu\sigma\Sigma=\frac{3}{2}\,\nu\,\Sigma\,\frac{M^{1/2}}{R^{3/2}}\,\frac{{\cal D}}{{\cal C}},
\end{equation}
where $\nu$ is the same effective viscosity that appeared in the
 non-relativistic expressions.  Noting that the total {\it torque} is
 given by $G=2\pi WR^2{\cal D}$, we see that the appropriate
 relativistic diffusion equation describing a sporadically torqued
 disk (i.e., the counterpart to eqn.~\ref{eqn:diffusion}) is
\begin{equation}
\label{eqn:rel_diffusion}
\frac{\partial\Sigma}{{\partial}t}=\frac{{\cal C}^{1/2}}{2\pi {\cal B}R}\frac{\partial}{\partial R}\left[\frac{\Gamma}{\frac{\partial L^\dagger}{\partial R}}\frac{\partial}{\partial R}\left(\frac{3{\cal D}^2}{2{\cal C}}\nu\Sigma M^{1/2}R^{1/2}-\frac{G_T}{2\pi}\right)\right].
\end{equation}
It is straightforward to verify that eqn.~\ref{eqn:rel_diffusion}
reduces to eqn.~\ref{eqn:basic_diffusion} in the non-relativistic
limit (i.e., ${\cal B}, {\cal C}, {\cal D}\rightarrow 1;
L^\dagger\rightarrow (MR)^{1/2}$).  The complications introduced by
the relativistic factors render this equation intractable to
elementary solution methods.  Thus, we use a simple explicit scheme to
numerically solve this diffusion equation following the treatment of
Press et al. (1992).  Figures~\ref{fig:relativistic} and 4 show the
temporal behavior of an accretion disk around Schwarzschild ($a=0$)
and near maximal Kerr ($a=0.998$) black holes respectively.  To
facilitate comparison with the non-relativistic case, we have chosen
the same viscosity law, $\nu=kR$ (i.e. $\nu$ scales with the radial
Boyer-Lindquist coordinate and not the proper distance).  As in the
Newtonian case, this prescription does not change the results
qualitatively.  Note that the behavior is similar to that found in the
non-relativistic model.  The two phases of evolution, the damming
phase and the replenishing phase, are reproduced.  The differences do
not come from the dynamics but from the boundary conditions that are
determined by the equilibrium conditions for circular geodesics.  The
presence of an innermost stable circular orbit for general
relativistic potentials suggests placing the disk inner boundary at
$R=6M$ for Schwarszchild spacetime and $R=1.23M$ for the near-maximal
Kerr spacetime.

Relating the observed flux to the fundamental disk structure is
substantially more complex in the relativistic case due to the
complexities of general relativistic photon propagation.  For a given
 value of the stress $W$, energy conservation gives that the total
radiative flux from one side of the disk, measured in the locally
orbiting frame, is (Novikov \& Thorne 1974),
\begin{equation}
F(R)=\frac{3{\cal D}}{4{\cal C}}\Omega W.
\end{equation}
Suppose that the corresponding energy-integrated (but angle-dependent)
intensity is $I_e(R, \theta)$, where $\theta$ is measured from the
normal to the disk plane (in the locally orbiting frame of reference).
Following Cunningham (1975), an observer at infinity will see an
integrated luminosity,
\begin{equation}
L_0=\int\int 2\pi I_e \Upsilon g^3 (g^*-g^{*2})^{-1/2}\,dg^*\,d(\pi R^2)
\end{equation}
where we have followed the notation of Cunningham (1975) with the
exception of $\Upsilon$.  Here $\Upsilon$ is the relativistic transfer
function that results from ray-tracing null-geodesics through the
Kerr metric from the disk to the observer.  We have used the code of
R.Speith (Speith, Riffert \& Ruder 1995) to compute $\Upsilon$ and
hence perform this integral in order to examine how the observed
luminosity changes through the torquing event.

These calculations uncover a fundamental difference between the
Newtonian and relativistic cases. In the relativistic case, the
observed changes in luminosity are a function of the inclination angle
of the observer due to the effects of light bending and relativistic
beaming of the disk emission.  Figures~3f and 4f show the temporal
behaviour of the observed flux (normalized to the flux for the
untorqued disk) for various observing angles for our Schwarzschild and
near-extremal Kerr cases, respectively.  The final fractional increase
in observed flux depends on the beaming pattern of the torque-energize
region of the disk compared with that of the untorqued disk.  For our
Schwarzschild case (Fig.~3f), one can see that the final fractional
increase in observed luminosity depends very weakly on the observing
angle, implying that the untorqued disk and the torque-energerized
region of the torqued disk have very similar beaming patterns.  There
is, however, a much more pronounced temporary decrease in observed
flux at higher inclinations due to the temporary dimming of the (more
highly beamed) inner regions of the disk.  For our Kerr case, the
final fractional increase in observed luminosity increases by almost a
factor of two as one moves from almost face-on to almost edge-on
disks, implying that the torque-energized region of the disk is
significantly more beamed than the untorqued disk.  A temporary
decrease is only observed for the most edge-on cases, again due to a
temporary dimming of the inner most regions of the disk.

These features are all symptomatic of the fact that our black hole
torques our accretion disk and deposits energy and angular momentum in
the disk.  This extra source of energy that is dumped into the disk
and that tends to affect the disk outward of the external torque
location constitutes the starting point for the analysis of Section 4
where we attempt to explain the "deep minimum state" as the result of
just such a sporadic torquing event.

\begin{figure*}
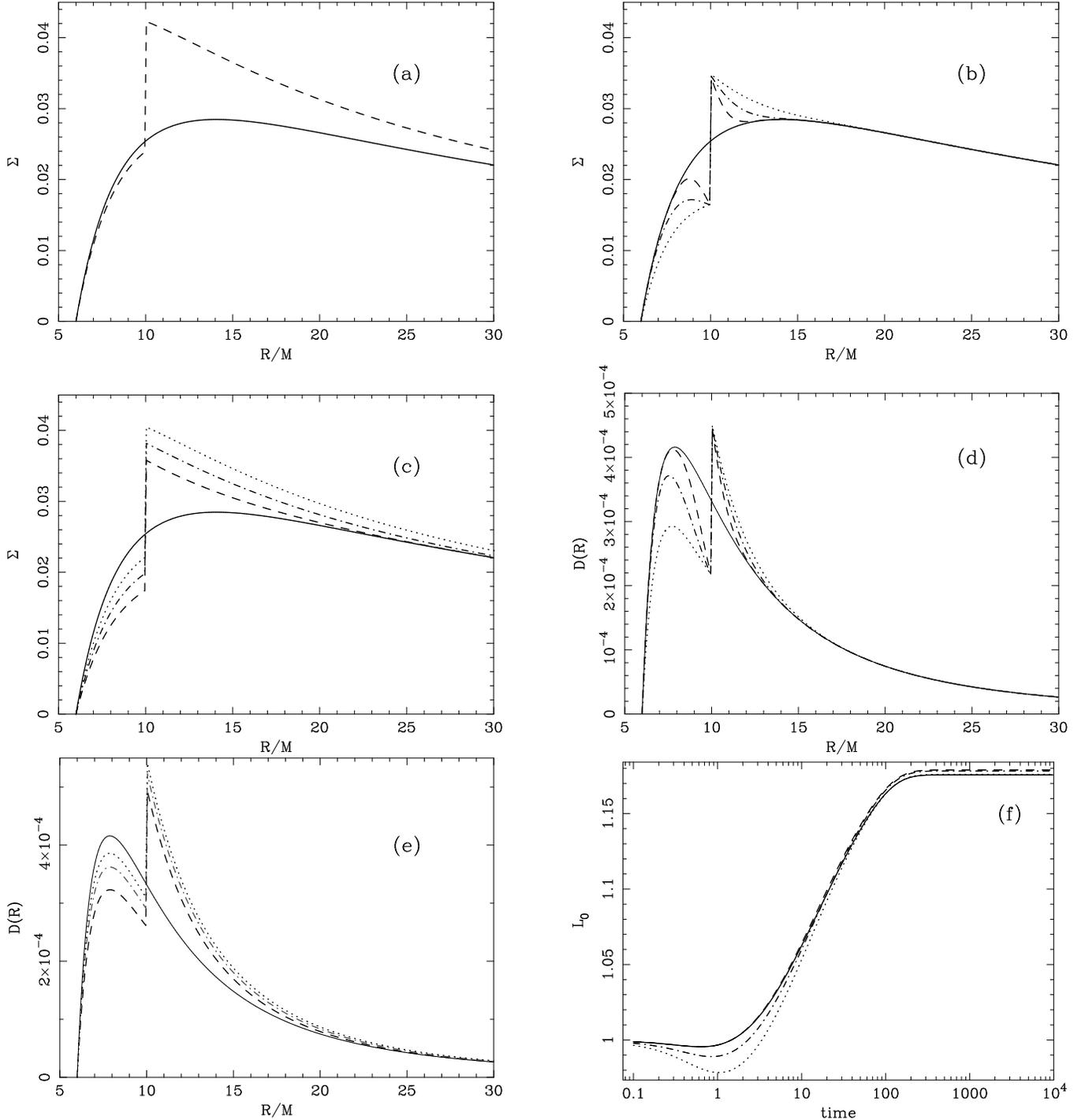

\hbox{
\psfig{figure=f3a.ps,width=0.45\textwidth,angle=270}
\hspace{1cm}
\psfig{figure=f3b.ps,width=0.45\textwidth,angle=270}
}
\hbox{
\psfig{figure=f3c.ps,width=0.45\textwidth,angle=270}
\hspace{1cm}
\psfig{figure=f3d.ps,width=0.45\textwidth,angle=270}
}
\hbox{
\psfig{figure=f3e.ps,width=0.45\textwidth,angle=270}
\hspace{1cm}
\psfig{figure=f3f_new.ps,width=0.45\textwidth,angle=270}
}
\label{fig:relativistic}
\caption{Evolution of disk in Schwarzschild spacetime for torque at R/M=10.  
  Panel (a) shows the surface density profile just after the torquing
  event begins as well as the steady-state torqued profile is
  approached (dashed-line:$t=10000$).  Panel (b) shows the early
  stages in the evolution of the surface density profile with the
  solid line being the untorqued steady-state profile
  (dashed-line:$t=0.8$, dot-dashed-line:$t=2.53$,
  dotted-line:$t=8.0$).  Panel (c) shows the untorqued steady-state
  profile (solid-line) as well as the late-time evolution of the
  torqued profile (dashed-line:$t=25$, dot-dashed-line:$t=80$,
  dotted-line:$t=253$).  Panel (d) shows the early evolution of the
  dissipation function with lines and times corresponding to those of
  panel (b).  The qualitative feature is again a drop inward of the
  torque location and an increase outward.  Panel (e) shows the
  late-time evolution of dissipation function with lines and times
  analogous to those of panel (c) while panel (f) shows the observed
  luminosity starting at untorqued steady-state with t=0.  The
  observed luminosity is determined for angles of 10 (solid-line), 30
  (dashed-line), 60 (dot-dashed line) degrees, and 80 degrees (dotted
  line).  Although the magnitude of the observed luminosity is not the
  same in the untorqued steady-state for all angles, we have
  normalized them in order to see the change with respect to the
  untorqued state.  Note the presence of a drop in the luminosity as
  the angle of inclination decreases.  }
\end{figure*}

\begin{figure*}
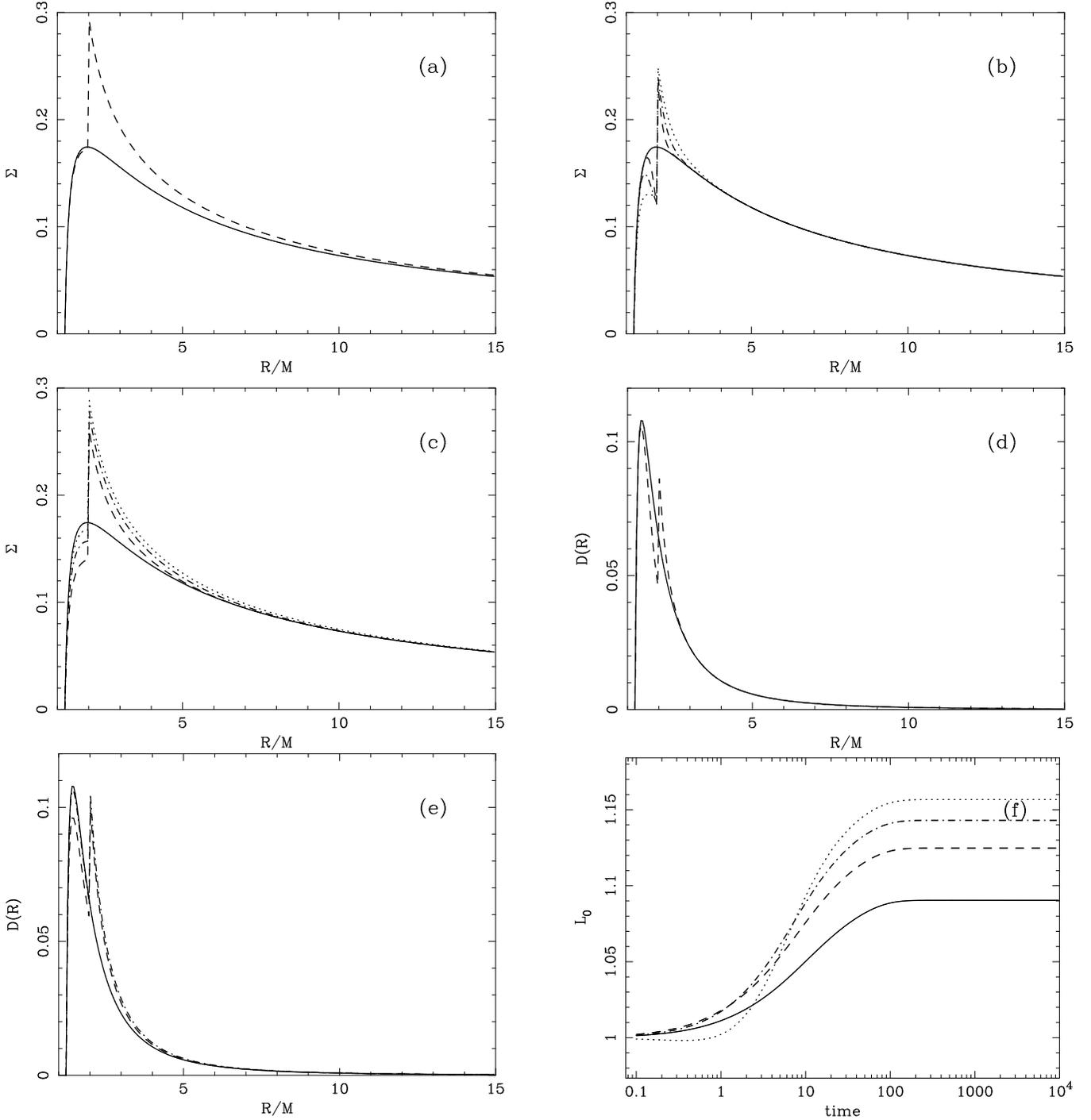

\hbox{
\psfig{figure=f4a.ps,width=0.45\textwidth,angle=270}
\hspace{1cm}
\psfig{figure=f4b.ps,width=0.45\textwidth,angle=270}
}
\hbox{
\psfig{figure=f4c.ps,width=0.45\textwidth,angle=270}
\hspace{1cm}
\psfig{figure=f4d.ps,width=0.45\textwidth,angle=270}
}
\hbox{
\psfig{figure=f4e.ps,width=0.45\textwidth,angle=270}
\hspace{1cm}
\psfig{figure=f4f_new.ps,width=0.45\textwidth,angle=270}
}
\label{fig:relativistic}
\caption{Evolution of disk in Kerr spacetime for torque at R/M=2 and 
  spin parameter $a=0.998$.  Panel (a) shows the surface density
  profile in untorqued steady-state (solid-line) as well as the
  approach to steady-state torqued profile (dashed-line:$t=10000$).
  Panel (b) shows the early stages in the evolution of the surface
  density profile with the solid line being the untorqued steady-state
  profile (solid-line:$t=0$) and the other profiles matching the times
  and line styles for panel (b) of the Schwarzschild figure.  Panel
  (c) shows the untorqued steady-state profile (solid-line) as well as
  the late-time evolution of the torqued profile (dashed-line:$t=25$,
  dot-dashed-line:$t=80$, dotted-line:$t=253$).  Panel (d) shows the
  early evolution of the dissipation profile in addition to the
  untorqued steady-state (solid-line) for the same times and line
  styles of the Schwarzschild panel (d).  Panel (e) shows the late
  stage evolution of the dissipation function with times and
  line-styles compatible with those of figure 3e.  Panel (f) shows the
  luminosity observed at the same angles as in the Schwarzschild case
  (10 degrees, solid-line; 30 degrees, dashed-line; 60 degrees,
  dotted-dashed-line; 80 degrees, dotted line).  Note how the smallest
  torqued steady-state rise occurs for the intermediate angle of 55
  degrees.  The lack of a drop in the observed luminosity comes from
  the presence of the external torque nearer to the inner boundary in
  the radial coordinate than in the Schwarzschild case.  }
\end{figure*}

\section{Can we interpret the ``Deep Minimum State'' of MCG--6-30-15 
as a sporadic torquing event?}

In addition to exploring the general characteristics of
sporadically-torqued disks, a central motivation for this study are
the recent {\it XMM-Newton} observations of the Seyfert galaxy
MCG--6-30-15.  In particular, we would like to explore whether the
enigmatic ``Deep Minimum State'' of this AGN could correspond to a
sporadic torquing event, possibly induced by the formation of a
temporary magnetic connection between the inner accretion disk and
either the plunging region of the disk or the rotating event horizon.
There are two defining characteristics of the Deep Minimum State that
must be reproduced by any successful model, the extremely broadened
X-ray reflection features (implying a very centrally concentrated
X-ray irradiation pattern) and the factor 2--3 drop in the observed
X-ray continuum flux.

A major uncertainty when relating disk models to X-ray observations is
always the relation between the dissipation within the disk (predicted
by the models) and the emission of the observed X-rays.  If we suppose
that a local disk-corona radiates a fixed fraction of the underlying
dissipation into the X-ray band, the results of this paper quickly
lead to a contradiction between the sporadically-torqued disk model
and the observations.  While the model does predict a temporary dip in
observed luminosity for some observer inclinations (that one might be
tempted to identify with the continuum drop in the Deep Minimum), this
dip is due to a dimming of the innermost regions of the accretion flow
as a result of the damming of the mass flux.  This is precisely the
part of the flow that we wish to be enhanced in order to explain the
simultaneous broadening of the X-ray reflection features.

With the (standard) accretion disk corona framwork, the relation
between the dissipation within the disk and the emission of the
observed X-rays depends on the reprocessing of disk photons by the
corona. We suggest this relation changes when the system departs from
steady-state as the sporadic torque engages and will use this, in the
next section, to model the Deep Minimum spectrum.

\subsection{Quenching the X-ray corona with returning radiation}

The assumption that the X-ray emission from the disk corona locally
tracks the dissipation in the underlying accretion disk is clearly an
oversimplication.  For example, Krolik \& Hawley (2001) have used
high-resolution pseudo-Newtonian simulations to show that there is a
rather extended transition (occurring near but slightly outside of the
radius of marginal stability) from the pure MHD turbulent region
characterizing the bulk of the disk to the more laminar flow present
in the plunging region.  Since the heating of the corona is almost
certainly due to reconnection and MHD wave heating from the underlying
disk, the fraction of the dissipated energy transported to the corona
will certainly change within this transition region, leading to a
violation of the simple assumption employed in our toy models.  A
time-variable magnetic torque of the kind we envision in this paper
might alter the MHD and thermodynamic properties of the gas and such a
scenario might not be compatible with the one we describe in the
thin-disk approximation.  In other words, we can imagine that the
external torque changes both the radiative efficiency of the gas as
well as local MHD properties thereby invalidating the treatment of
the magnetorotational instability as a local kinematic viscosity.
This fact is most likely more important for thick disks where the
degrees of freedom are greater.  Global disk simulations focusing on
the formation and properties of the corona are required to address
this issue and, hence, are beyond the scope of this paper.

We do, however, note an important and mostly neglected physical effect
that could substantially change the structure of a disk corona in a
strongly torqued disk --- Compton cooling by flux emitted elsewhere in
the accretion disk and, in particular, by ``Returning Radiation''.
Consider a geometrically-thin accretion disk around a near-extremal
Kerr black hole, and suppose that it possesses a disk-hugging X-ray
corona energized from the underlying disk.  Now suppose that the
central regions of the disk are subjected to a significant torquing
event.  As shown above, the work done by the torque is rapidly
radiated from the accretion disk in a very centrally concentrated
manner.  The torque-induced emission will be a combination of both
thermal optical/UV radiation and hard X-ray emission produced by the
corona associated with the torque-energized regions of the disk.  Now,
some fraction of the torque-induced emission will strike the disk at
larger radii --- this will be particularly prevalent if the disk is
flared or warped, but will occur even in flat disks due to
relativistic light bending effects (i.e., Returning Radiation;
Cunningham 1973).  This extra irradiation will enhance the Compton
cooling of the corona at these larger radii.  At the very least, the
additional cooling will lead to a decrease in the Compton
amplification factor of the corona and a steepening of the coronal
emission.  One could envisage a situation, however, in which the
Compton cooling becomes so extreme that the corona completely
collapses and local EUV/X-ray emission ceases.

Some essential aspects of this scenario can be captured in a simple
model based on energy conservation, following Haardt \& Maraschi
(1991, 1993).  Consider the X-ray emitting corona above a unit-area
patch of the disk at a radius $r$.  If a fraction $f$ of the energy
dissipated in the underlying disk goes into heating the corona, the
heating rate is
\begin{equation}
{\cal H}(r)=fD(r).
\end{equation}
By definition, the (Compton) cooling rate of the corona is
$(A-1)F_{\rm s}$, where $A$ is the Compton amplification factor and
$F_{\rm s}$ is the soft photon flux passing through the corona which
will act as seed photons for the inverse Compton scattering process
that generates the X-rays.  Equating heating and cooling gives,
\begin{equation}
fD(r)=(A-1)F_{\rm s}.
\end{equation}
We now determine $F_{\rm s}$ by examining energy conservation of the
colder disk underlying the corona.  There are three contributions to
that we must consider.  Firstly, the portion of the internal
dissipation within the disk that does {\it not} get transported into
the disk will become thermalized in the cold disk and contribute an
amount $(1-f)D(r)$ to the disk heating.  Secondly, some fraction of
the locally generated coronal flux $\xi_1 fD(r)$ will impinge on the
disk and be reprocessed into soft flux.  The parameter $\xi_1$
encapsulates possible anisotropies in the coronal flux and the albedo
of the disk, but will typically be of the order of $\xi_1\sim
0.2-0.5$.  Finally, as noted above, irradiation of our coronal patch
from other radii in the disk will contribute to the soft flux and
hence the Compton cooling.  This will cool the corona due to both the
direct action of the irradiating soft flux, and the
reprocessing/thermalization of the soft and hard irradiating flux.
Suppose that the non-local irradiating flux is ${\cal R}(r)$ times the
locally produced flux.  The corresponding soft flux contributing to
the Compton cooling will be $\xi_2{\cal R}D(r)$, where $\xi_2\approxlt
1$ parameterizes the fraction of this non-local emission that ends up
as soft flux.  Hence, the total soft flux at a particular location in
the disk will be
\begin{equation}
F_{\rm s}=\xi_1fD(r)+\xi_2{\cal R}(r)D(r)+(1-f)D(r).
\end{equation}
Solving for $A$, we get
\begin{equation}
A=1+\frac{f}{\xi_2{\cal R}(r)+1-f(1-\xi_1)}.
\end{equation}

Of course, within this simple model the total energy dissipated within
the corona is a fixed fraction of the underlying dissipation
irrespective of the (cooling) soft flux.  However, the amplication
factor is significantly reduced by Returning Radtion if ${\cal
  R}(r)\approxgt \xi_2^{-1}[1-f(1-\xi_1)]$ which, for canonical values
of $f=1$ and $\xi_1=\xi_2=0.5$, corresponds to ${\cal R}(r)\approxgt
1$.  The resulting coronal spectrum from the affected regions of the
disk would be expected to steepen significantly, possibly placing a
large fraction of the emission into the unobservable EUV band.

For a flat disk at large radii subjected to returning radiation, we
have ${\cal R}(r)=R_0(a)+\Delta \eta R_{\infty}(a)$, where
$\Delta\eta$ is the enhancement in the efficiency of the disk due to
the inner torque and $R_{\infty}(a)$ and $R_0(a)$ are dimensionless
functions of the black hole spin parameter given by the fitting
formulae of Agol \& Krolik (2001).  For a near-extremal Kerr black
hole ($a=0.998$), we have $R_0\approx 0.2$ and $R_\infty\approx 1$.
Thus, we can see that even in the absence of disk flaring or warping,
returning radiation alone could significantly depress coronal X-ray
activity at large radii if
\begin{equation}
\Delta\eta\approxgt \frac{1-f[1-\xi_1]}{\xi_2}-0.2.
\end{equation}
Thus, although there is some dependence on the properties (e.g.,
isotropy and patchiness) of the corona and the ability of the disk to
reprocess and thermalize any incoming flux, the corona will be
depressed if the disk is in a ``spin-dominated'' state ($\Delta\eta
\approxgt 1$), i.e., a state in which the disk is shining via the
release of black hole spin energy rather than gravitational potential
energy.

\subsection{A proposed scenario for the MCG-6-30-15 Deep Minimum State}

Let us now return to MCG--6-30-15 and the sporadic external torque
model for its Deep Minimum State.  We suppose that the normal state of
this system is that of a standard untorqued accretion disk that might
well be described by the standard accretion models of Novikov \&
Thorne (1974) and Page \& Thorne (1974).  We then suppose that some
shift in magnetic configuration caused the accretion disk to become
magnetically torqued by either the plunging region or the rotating
black hole itself.  We hypothesize that this event signals the onset
of a Deep Minimum State.  

On timescales shorter than the viscous timescale of the inner disk
($t_{\rm visc}\sim 1\,{\rm hour}$), we expect this torquing event to
lead to a damming of the accretion flow and a true dimming of the disk
interior to the location where the connection has occurred.  However,
on longer timescales, the disk will tend to the new torqued
steady-state (provided the torque is sufficiently long-lived).  

If the magnetic torquing occurs in the very centralmost regions of the
disk (which is likely in all of the scenarios that we are envisaging),
the torqued steady-state will possess a much more centrally
concentrated dissipation pattern.  As described above, some fraction
of this central flux will strike the disk further out (the Returning
Radiation phenomenon) and possibly lead to a Compton suppression of
the X-ray emitting corona there.  Only the central portions of the
X-ray emitting corona which are being vigorously energized will
contributed significantly to the observed X-ray flux.  

It is simple to see that the {\it overall X-ray luminosity} escaping
the system is unlikely to drop, and will probably rise, within this
scenario --- the non-local cooling is only important for
$\Delta\eta>1$, in which case the part of the disk directly energized
by the torque will produce coronal luminosity in excess of
$f\dot{M}c^2$.  Even accounting for the fact that half of this may
strike the disk and be reprocessed into soft flux, the {\it overall}
X-ray luminosity of the torqued disk will inevitably exceed that of
the untorqued disk ($0.3\dot{M}c^2$).  However, the highly centrally
concentrated nature of the torqued emission coupled with the
suppression of the X-ray emission at larger radii means that this
X-ray luminosity is highly beamed into the plane of the disk (see
Fig.~10 of Dabrowski et al.  1997), and the {\it observed X-ray flux}
for an observer with an inclination of 30$^\circ$ can readily drop.
In other words, the observed coronal activity is supressed by cooling
from a powerful photon source that goes largely unobserved due to
beaming effects.

\section{Discussion and conclusions}

Both non-relativistic and fully-relativistic MHD simulations of black
hole accretion suggest the ubiquity of weak torques across the radius
of marginal stability (Hawley \& Krolik 2000, 2001; De~Villers, Hawley
\& Krolik 2003; Gammie, Shapiro \& McKinney 2004).  However, it is
still far from clear whether these torques can ever achieve the
strength required to allow a decline of the overall mass-energy of the
black hole, due to spin-down of the hole, a requirement
for the kind of spin-dominated accretion disks that we have been
discussing within the context of MCG--6-30-15.  Significantly more
simulation work is needed to address this question.

If one makes the assumption that such strongly torqued disks are
possible, they provide a plausible and theoretically-attractive
explanation for the extremely broadened X-ray reflection features seen
in the Deep Minimum State of MCG--6-30-15 (Wilms et al. 2001; Reynolds
et al. 2004) as well as the Galactic Black Hole Binary XTEJ1650--500
(Miller et al. 2002).  Given this hypothesis, we can ask whether the
onset of the Deep Minimum State corresponds to the occurrence of a
strong torquing event.  

We address this question through analytic and semi-analytic
time-dependent toy models of torqued accretion.  Using these models,
we have shown that the simple model of a sporadically-torqued disk
{\it fails} to explain the phenomenology of the Deep Minimum State
transition.  However, it would be premature to dismiss torqued-disk
models on the basis of this failure.  In particular, the real culprit
may be the assumption that the local X-ray emission is a fixed fraction of
the dissipation in the underlying disk.  We discuss a particular
scenario in which all but the innermost X-ray emitting corona is
quenched by returning radiation when the torque is engaged.  Other
possibilities include changes in the structure of the MHD turbulence
as a result of the strong torquing event that, in turn, could readily
change the fraction of the dissipated energy that is transported into
the corona.

An alternative paradigm is the gravitational light-bending model of
Fabian \& Vaughan (2003) and Miniutti \& Fabian (2004).  Here, the
primary X-ray source is located on the black hole spin axis.  A
transition into the Deep Minimum State corresponds to a migration of
the X-ray source down to 2--3\,$GM/c^2$, with the light bending
producing both an enhancement in the central illumination of the
accretion disk and a dimming of the observed X-ray continuum flux.
While the physical nature of the axial X-ray source remains unclear,
it is an appealing aspect of this model that it reproduces the
long-term temporal behavior of the iron line strength (Miniutti \&
Fabian 2004).

As is often the case, it is uncertainty in the geometry of the primary
X-ray source that prevents us from distinguishing between the
torqued-disk and light-bending models.  The most promising
observational approach is to search for a reverberation delay between
short timescale flickering in the X-ray continuum and the
corresponding response in the X-ray reflection signatures.  If
measurements of this time-delay demonstrate that the X-ray source in
the Deep Minimum is indeed 2--3\,$GM/c^2$ (corresponding to
$10-15\,(M/10^6\Msun)$ light seconds) above the central disk plane,
strong light-bending must occur and the need for a torqued disk is
removed.  Measurements of this time-delay will constrain the X-ray
source geometry and allow the degeneracy between these two models to
be resolved.

\section*{Acknowledgments}

We thank Andy Fabian, Ted Jacobson, Julian Krolik, Barry McKernan,
Cole Miller, Eve Ostriker, Brian Punsly, Joern Wilms and Andy Young
for stimulating conversations while this work was being performed, and
Laura Brenneman whose code provided the basis for the generation of
figures 3f and 4f.  We also thank the anonymous referee for useful
comments that lead to the improvement of this paper.  We gratefully
acknowledge support from the National Science Foundation under grant
AST0205990.

\section*{References}

\noindent Agol, E., Krolik, J.H., 2000, ApJ, 528, 161.

\noindent Armitage P.J., Livio M., Pringle J.E., 1996, ApJ, 457, 332.

\noindent Blandford, R.D., Znajek, R.L., 1977, MNRAS, 179, 433.

\noindent Cunningham, C.T., 1973, PhD Thesis, Univ. of Washington.

\noindent Cunningham, C.T., 1975, ApJ, 202, 788.

\noindent De Villiers, J.P., Hawley, J.F., Krolik, J.H., 2003, ApJ, 599, 1238.

\noindent Eardley D., Lightman A., 1975, ApJ, 200, 187. 

\noindent Fabian A.C., et al., 1989, MNRAS, 238, 729.

\noindent Fabian A.C., et al., 1995, MNRAS, 277, L11.

\noindent Fabian A.C., Vaughan, S., 2003, MNRAS, 340, L28.

\noindent Ford R.J., et al., 1994, ApJ, 435, L27.

\noindent Frank, J., King, A., Raine, D., 2002, in Accretion Power in Astrophysics, Cambridge Univ. Press, Cambridge.

\noindent Gammie, C.F., 1999, ApJ, 522, 57.

\noindent  Gammie, C.F., Shapiro, S.L., McKinney, J.C., 2004, ApJ, 602, 312.

\noindent Greenhill, L.J., 1995, ApJ, 440, 619.

\noindent Harms, H.C., et al., 1994, ApJ, 435, L35.

\noindent Hawley, J.F., Krolik, J.H., 2001, ApJ, 548, 348.

\noindent Hawley, J.F., Krolik, J.H., 2002, ApJ, 566, 164.

\noindent Hirose, S., et al., 2004, ApJ, 606, 1083.

\noindent Iwasawa, K., et al., 1996, MNRAS, 282, 1038.

\noindent Krolik J.H., 1999, ApJ, 515, L73.

\noindent Krolik J.H., 1999, in Active Galactic Nuclei, Princeton University Press, Princeton, 139.

\noindent Li, Li-Xin, 2002, ApJ, 567, 463.

\noindent Lightman A., 1974, ApJ, 194, 419.

\noindent Miller, J.M., et al., 2002, AAS, 34, 1206.

\noindent Miniutti, G., Fabian, A.C., 2004, MNRAS, 349, 1435.

\noindent Miyoshi et al., 1995, Nature, 373,127

\noindent Novikov, I., Thorne K.S., 1973, in Black Holes, 1973, ed. C.DeWitt and B.DeWitt(New York: Gordon and Breach).

\noindent Page, D.N., Thorne, K.S., 1974, ApJ, 191, 499.

\noindent Press, W.H., et al., 1992, Numerical Recipes in C. The Art of Scientific Computing (Cambridge Univ. Press).

\noindent Pringle J.E., 1981, ARAA, 19, 137

\noindent Reynolds C.~S., Nowak M.A., 2003, Physics Reports, 377,389.

\noindent Reynolds C.S., Wilms J., Begelman M.C., Staubert R., Kendziorra E., 
2004, MNRAS, 349, 1153.

\noindent Riffert H., Herold H., 1995, ApJ, 450, 508.

\noindent Shakura, N.I., Sunyaev, R.A., 1973, A\&A, 24,337. 

\noindent Speith, R., Riffert, H., Ruder, H., 1995, Computer Physics Communications, 88, 109. 

\noindent Tanaka Y., et al., 1995, Nature, 375.

\noindent Vaughan, S., Fabian, A.C., Nandra, K., 2003, MNRAS, 339, 1237.

\noindent Williams R.,K., 2003, astro-ph/0306135 v1

\noindent Wilms J., et al., 2001, MNRAS, 328, L27

\end{document}